\documentclass[fleqn,10pt]{wlscirep}
\usepackage{etex}
\usepackage[utf8]{inputenc}
\usepackage[T1]{fontenc}
\usepackage{multirow}
\usepackage{tikz}
\usepackage{pgfplots}
\usepackage [lined,boxed,commentsnumbered]{algorithm2e}
\usepackage{subcaption}
\usepackage{adjustbox}
\usepackage{floatrow}

\usetikzlibrary{decorations.pathreplacing}
\pgfplotsset{width=8cm,compat=1.5}

\title{Fast simulation of Grover's quantum search on classical computer}

\author[1,*]{Ayan Chattopadhyay}
\author[2]{Vikram Menon}
\affil[1,2]{Independent Researcher}

\affil[*]{ayan.eab@gmail.com}

\begin{abstract}
We provide a classical approach for the fast simulation of the Grover's quantum search algorithm as compared to the existing known quantum simulations. The topic is important since Grover's search algorithm is compute intensive and has wide use in various quantum computing applications. The existing simulators implement the quantum circuit using universal gates, while the proposed method improves the number of calculations by realizing an equivalent mathematical operator, resulting in a significant improvement in the running time. This will enable researchers to quickly verify their computations on a personal computer. Such a  simulator will be helpful until real quantum computers are a reality and readily accessible.

\end{abstract}

\begin{document}

\flushbottom
\maketitle

\thispagestyle{empty}

\section*{Introduction}
Grover's iterative quantum search algorithm \cite{grov} can be used in such problems as cryptography, AI, pattern matching, database search, and is the most efficient quantum search algorithm till date. Its algorithmic complexity is $O(\sqrt{N})$, where $N$ is the size of the search space. Grover's quantum iterator has the following components (1) oracle for selected state inversion, (2) Hadamard transformation, (3) conditional phase shift to all the states except $|0\rangle$, and (4) Hadamard transformation. Our purpose here is to provide efficient simulation of above mentioned quantum operations on a classical Turing Machine (TM).
In general, simple minded classical simulation of quantum algorithms have an obvious problem - parallel quantum operations on superposed quantum states must be serialized on a TM or parallelized on multiple TMs. The former leads to execution time growing exponentially, while the latter leads to physical resources growing exponentially. Classical simulation therefore quickly becomes in-feasible. In this report, we outline classical implementation method that significantly reduces the time-space resource requirements for simulating Grover's algorithm on classical TM.

Conventional quantum simulators \cite{quest},\cite{libquantum} focus on simulating quantum circuit using universal gates, which can be executed on a quantum computer. For any circuit the number of gates increases with the number of qubits and the simulation of every unitary transformation requires a loop of size $N = 2^n$, where $n$ is the number of qubits. The Grover's search algorithm, due to the iterative nature, is therefore the most compute intensive. However, the performance (both run-time and performance) degrades with number of qubits in the system or the search space. We will show that the run-time can be improved significantly by implementing the equivalent mathematical circuit, without affecting the outcomes. The objective is to make the simulation computationally cheaper and time-efficient by minimizing the number of transformations and therefore loops, thus making it independent of the search space. 

\section{Background}
The Grover search algorithm has four stages:
\begin{enumerate}
\item Apply Hadamard transform $H^{\otimes n}$ to create equal superposition of all the states.
\item Apply Oracle $O$ to flip the solution state(s).
\item Apply Hadamard transform $H^{\otimes n}$.
\item Perform Conditional Phase Shift $2|0\rangle \langle 0| - I$ to flip all the states except $|0\rangle^{\otimes n}$.
\item Apply Hadamard transform $H^{\otimes n}$.
\item Perform Measurement.
\end{enumerate}
Steps $3-5$ constitute the inversion about the mean or amplitude amplification operator $2|\psi\rangle \langle \psi| - I$ and $2-5$, the Grover iterator $G=(2|\psi\rangle \langle \psi| - I)O$.

The following subsections outline the reference implementation of the Grover's search on a conventional quantum simulator, as a basis to compare the proposed method defined in the subsequent section.

\subsection{Quantum State}
A {\em quantum state} can be realized using a linear array of complex numbers, where the indices represent the basis states and the values map to the associated probability amplitudes. The representation of a generic $n$-qubit system state $\psi = \sum_{i=0}^{N-1} c_{i} |i\rangle$ is shown below,

\begin{table}[!ht]
  \begin{center}
    \begin{tabular}{|c|c|c|c|c|c|c|c|c|}
      \hline
      \multirow{2}{*}{$\psi[N]$} & Basis State $\rightarrow$ & 0 & 1 & 2 & 3 & $\cdots$ & N-2 & N-1 \\
      \cline{2-9}
       & Probability Amplitude $\rightarrow$ &  $c_{0}$ & $c_{1}$ & $c_{2}$ & $c_{3}$ & $\cdots$ & $c_{N-2}$ & $c_{N-1}$ \\
      \hline
    \end{tabular}
  \end{center}
\end{table}
where $c_{i}$ represents the probability amplitude of the basis state $i$, $\sum_{i} |c_{i}|^{2} = 1$ and $N=2^{n}$.

\subsection{Unitary Gate}
A {\em single qubit unitary gate} can be represented by a $2X2$ unitary matrix, $U = \begin{pmatrix} a_{11} & a_{12} \\ a_{21} & a_{22}\end{pmatrix}$, with $UU^{\dag} = I$. Equation \ref{eqn:unitary} describes the transformation applied to a generic qubit state.
\begin{equation}
\label{eqn:unitary}
\begin{split}
  U_{1}|0\rangle &= \begin{pmatrix} a_{11} & a_{12} \\ a_{21} & a_{22} \end{pmatrix} \begin{pmatrix} 1 \\ 0\end{pmatrix} = \begin{pmatrix} a_{11} \\ a_{21} \end{pmatrix} = (a_{11}|0\rangle + a_{21}|1\rangle)\\
  U_{1}|1\rangle &= \begin{pmatrix} a_{11} & a_{12} \\ a_{21} & a_{22} \end{pmatrix} \begin{pmatrix} 0 \\ 1\end{pmatrix} = \begin{pmatrix} a_{12} \\ a_{22} \end{pmatrix} = (a_{12}|0\rangle + a_{22}|1\rangle)\\
  U_{1}(\alpha |0\rangle + \beta |1\rangle) &= (\alpha U_{1}|0\rangle + \beta U_{1}|1\rangle) \\
  &= \begin{pmatrix} \alpha * a_{11} \\ \alpha * a_{21} \end{pmatrix} + \begin{pmatrix} \beta * a_{12} \\ \beta * a_{22} \end{pmatrix} = \begin{pmatrix} \alpha * a_{11} + \beta * a_{12} \\ \alpha * a_{21} + \beta * a_{22} \end{pmatrix}\\
  &= (\alpha * a_{11} + \beta * a_{21})|0\rangle + (\alpha * a_{12} + \beta * a_{22})|1\rangle
\end{split}
\end{equation}
For a $n$ qubit arbitrary state, the operation of $U$ on $i^{th}$ target qubit can be simulated as follows. Apply the above transformation to the $i^{th}$ qubit of all the $N=2^{n}$ basis states. It is evident that the unitary operation on $i^{th}$ target qubit of a given basis state, say $|x_{n-1}x_{n-2}\dotsb x_{i} \dotsb x_{1} x_{0}\rangle$, will affect the probability amplitude of the basis state $|x_{n-1}x_{n-2} \dotsb \bar{x_{i}} \dotsb x_{1} x_{0}\rangle$, where $\bar{x_{i}} = \begin{pmatrix} 0 & 1 \\ 1 & 0 \end{pmatrix}x_{i}$. Algorithm \ref{alg:unitary} outlines a possible implementation.

\begin{algorithm}[H]
  \caption{1-qubit Unitary Operation}
  \label{alg:unitary}
  \SetAlgoLined
  \SetKwFunction{FMain}{U}
  \SetKwProg{Fn}{Function}{:}{end}
  \Fn{\FMain{state: $\psi[N]$, target qubit: $i$}}{
  \KwData{$U = \begin{pmatrix} a_{11} & a_{12} \\ a_{21} & a_{22}\end{pmatrix}$}
  \For {$basis\_state=0;\ basis\_state < N-1;\ basis\_state++$}{
    \If {$\psi[basis\_state]$ present}{
        $zero\_target\_state = basis\_state \& ~(1 << i)$ \tcp*{state with the target qubit $|0\rangle$}
        $one\_target\_state = basis\_state | (1 << i)$ \tcp*{state with the target qubit $|1\rangle$}
        \eIf (\tcp*[h]{target qubit = $|0\rangle$}){$!(basis\_state \& (1 << i))$} {
            \tcp{$U|0\rangle = a_{11}*|0\rangle + a_{21}*|1\rangle$}
            $a1 = a_{11} * \psi[basis\_state]$\;
            $a2 = a_{21} * \psi[basis\_state]$\;
        }(\tcp*[h]{target qubit = $|1\rangle$}){
            \tcp{$U|1\rangle = a_{12}*|0\rangle + a_{22}*|1\rangle$}
            $a1 = a_{12} * \psi[basis\_state]$\;
            $a2 = a_{22} * \psi[basis\_state]$\;
        }
        $\psi[zero\_target\_state] += a1$\;
        $\psi[one\_target\_state] += a2$\;
    }
  }
  return 0\;
  }
\end{algorithm}
The run-time complexity of this method, is therefore, O($N$). Similarly, a $n$-qubit transformation will require $n$-iterations of the above function $U$ with a complexity of O($nN$).

\subsection{Oracle} \label{oracle}
The {\em Oracle}, $O$, is required to mark the solution state. Its operation can be summarized as $O[\frac{1}{\sqrt{N}} \sum_{x=0}^{N-1} |x\rangle] \Rightarrow \frac{1}{\sqrt{N}} \sum_{x=0}^{N-1} (-1)^{f(x)}|x\rangle$, where $f(x)=1$, if $|x\rangle$ is the solution and $0$ otherwise.

\begin{figure}[ht!]
\centering
\begin{tikzpicture}
  \draw (.5,3.25) node {$|\psi\rangle$};
  \draw (.4,2.4) node {$\frac{|0\rangle - |1\rangle}{\sqrt{2}}$};
  \draw (.4,1.7) node {Solution state:};
  \draw (1,3.5)  -- (1.5,3.5);
  \draw (1,3)  -- (1.5,3);
  \draw (1,2.4)  -- (3.8,2.4);
  \draw (1.5,2.1) rectangle (3.5,3.8) node [midway=center] {};
  \draw (1.6,3.3) rectangle (2,3.7) node [midway=center] {$X$};
  \draw (1.6,2.8) rectangle (2,3.2) node [midway=center] {$X$};
  \draw (2.5,3.5) node [shape=circle,draw,fill=black,inner sep=0pt,minimum size=2mm]{}  -- (2.5,3.5);
  \draw (2.5,3) node [shape=circle,draw,fill=black,inner sep=0pt,minimum size=2mm]{}  -- (2.5,3);
  \draw (2.5,2.4) node [shape=circle,draw,fill=none,inner sep=0pt,minimum size=3mm]{}  -- (2.5,2.4);
  \draw (2.5,3.5)  -- (2.5,2.25);
  \draw (3,3.3) rectangle (3.4,3.7) node [midway=center] {$X$};
  \draw (3,2.8) rectangle (3.4,3.2) node [midway=center] {$X$};
  \draw (2,3.5)  -- (3,3.5);
  \draw (2,3)  -- (3,3);
  \draw (3.5,3.5)  -- (3.8,3.5);
  \draw (3.5,3)  -- (3.8,3);
  \draw (2.5,1.7) node {$|00\rangle$};

  \draw (4,3.5)  -- (4.5,3.5);
  \draw (4,3)  -- (6.8,3);
  \draw (4,2.4)  -- (6.8,2.4);
  \draw (4.5,2.1) rectangle (6.5,3.8) node [midway=center] {};
  \draw (4.6,3.3) rectangle (5,3.7) node [midway=center] {$X$};
  \draw (5.5,3.5) node [shape=circle,draw,fill=black,inner sep=0pt,minimum size=2mm]{}  -- (5.5,3.5);
  \draw (5.5,3) node [shape=circle,draw,fill=black,inner sep=0pt,minimum size=2mm]{}  -- (5.5,3);
  \draw (5.5,2.4) node [shape=circle,draw,fill=none,inner sep=0pt,minimum size=3mm]{}  -- (5.5,2.4);
  \draw (5.5,3.5)  -- (5.5,2.25);
  \draw (6,3.3) rectangle (6.4,3.7) node [midway=center] {$X$};
  \draw (5,3.5)  -- (6,3.5);
  \draw (5,3)  -- (6,3);
  \draw (6.5,3.5)  -- (6.8,3.5);
  \draw (5.5,1.7) node {$|01\rangle$};

  \draw (7,3.5)  -- (9.8,3.5);
  \draw (7,3)  -- (7.5,3);
  \draw (7,2.4)  -- (9.8,2.4);
  \draw (7.5,2.1) rectangle (9.5,3.8) node [midway=center] {};
  \draw (7.6,2.8) rectangle (8,3.2) node [midway=center] {$X$};
  \draw (8.5,3.5) node [shape=circle,draw,fill=black,inner sep=0pt,minimum size=2mm]{}  -- (8.5,3.5);
  \draw (8.5,3) node [shape=circle,draw,fill=black,inner sep=0pt,minimum size=2mm]{}  -- (8.5,3);
  \draw (8.5,2.4) node [shape=circle,draw,fill=none,inner sep=0pt,minimum size=3mm]{}  -- (8.5,2.4);
  \draw (8.5,3.5)  -- (8.5,2.25);
  \draw (9,2.8) rectangle (9.4,3.2) node [midway=center] {$X$};
  \draw (8,3)  -- (9,3);
  \draw (9.5,3.5)  -- (9.8,3.5);
  \draw (9.5,3)  -- (9.8,3);
  \draw (8.5,1.7) node {$|10\rangle$};

  \draw (10,3.5)  -- (12.8,3.5);
  \draw (10,3)  -- (12.8,3);
  \draw (10,2.4)  -- (12.8,2.4);
  \draw (10.5,2.1) rectangle (12.5,3.8) node [midway=center] {};
  \draw (11.5,3.5) node [shape=circle,draw,fill=black,inner sep=0pt,minimum size=2mm]{}  -- (11.5,3.5);
  \draw (11.5,3) node [shape=circle,draw,fill=black,inner sep=0pt,minimum size=2mm]{}  -- (11.5,3);
  \draw (11.5,2.4) node [shape=circle,draw,fill=none,inner sep=0pt,minimum size=3mm]{}  -- (11.5,2.4);
  \draw (11.5,3.5)  -- (11.5,2.25);
  \draw (11,3.5)  -- (12,3.5);
  \draw (11,3)  -- (12,3);
  \draw (11.5,1.7) node {$|11\rangle$};
\end{tikzpicture}
\caption{2-qubit conventional quantum Oracles} \label{fig:oracle1}
\end{figure}
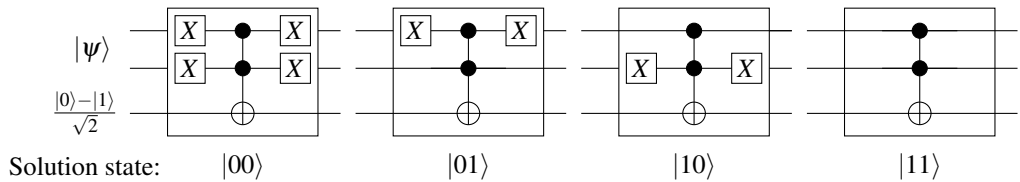

A possible $2$-qubit quantum oracle construction, for single solution, is captured in Fig. \ref{fig:oracle1}.

The run-time in this case is given by O($f_{o}(n)N$), where $f_{o}(n)$ is the number of universal gates required to build a $n$ qubit Oracle.
\subsection{Conditional Phase Shift}
Similar to the Oracle implementation, the number of gates required to build the conditional phase shift is $f_{c}(n)$. The complexity will then be O($f_{c}(n)N$).

\subsection{Simulation of Grover's Search}
\label{sec:qsim}
The circuit simulating the overall Grover search is captured in Fig. \ref{alg:grover1} below. 

\begin{algorithm}[H]
  \caption{$n$ qubit Grover's Search}
  \label{alg:grover1}
  \SetAlgoLined
  \SetKwFunction{FMain}{grover\_search}
  \SetKwProg{Fn}{Function}{:}{end}
  \Fn{\FMain{search state: $|i\rangle$}}{
  $\psi[N] = |0\rangle^{n}$\;
    $H^{\otimes n}(\psi[N])$ \tcp*{prepare superposition state}
   \tcp{Apply $\frac{\pi * \sqrt{N}}{4}$ iterations of Grover iterator}
  \For {$iter=0;\ iter < \frac{\pi * \sqrt{N}}{4};\ iter++$}{
      $oracle(\psi[N],i)$ \tcp*{Oracle: Flip the phase of solution state $|i\rangle$}
      $H^{\otimes n}(\psi[N])$ \tcp*{Apply Hadamard transformation}
      $conditional\_phase\_shift(\psi[N],0)$ \tcp*{Flip the phase of state $|0\rangle$}
      $H^{\otimes n}(\psi[N])$ \tcp*{Apply Hadamard transformation}
  }
  return 0\;
  }
\end{algorithm}

The overall time complexity is approximately O($\sqrt{N}*[f_{o}(n)N+nN+f_{c}(n)N+ nN]$) = O($[f_{o}(n)+f_{c}(n)+2n]N^{\frac{3}{2}}$).

\section{Proposed Simulation Method}
The Grover operator $G=(2|\psi\rangle \langle \psi| - I)O$ consists of two sub-operators, the Oracle ($O$) and the inversion about the mean operator ($2|\psi\rangle \langle \psi| - I$).
Instead of implementing the circuit using the universal gates as the building blocks, we propose to implement the equivalent mathematical operators that can be efficiently simulated on classical computers.  

\subsection{Oracle}
A typical classical realization of the Oracle construct, outlined in the previous section, is to flip the relative phase of the solution state. Since the Oracle is assumed to know the solution, this can be achieve in O($N$) instead of O($f_{o}(n)N$).

\begin{algorithm}[H]
  \caption{n-qubit Oracle}
  \label{alg:oracle}
  \SetAlgoLined
  \SetKwFunction{FMain}{oracle}
  \SetKwProg{Fn}{Function}{:}{end}
  \KwData{solution state(s): $k[]$}
  \Fn{\FMain{oracle: $|\psi\rangle$}}{
    \For {$i=0;\ i < sizeof(k);\ i++$}{
      $\psi[k[i]] *= -1$\;
    }
    return 0\;
  }
\end{algorithm}

\subsection{Inversion About Mean Operator}
The {\em inversion about mean operator} aka the {\em Diffusion operator}, for a given state $|\psi\rangle$, is given by $(2|\psi\rangle \langle \psi| - I)$ or $[H^{\otimes n}(2|0\rangle \langle 0| - I)H^{\otimes n}]$ and is mathematically equivalent to 
\begin{equation}
  \begin{split}
  \Delta_{i,j} &= \frac{2}{N} - 1; \forall i = j\\
  \Delta_{i,j} &= \frac{2}{N}; i \neq j
  \end{split}
\end{equation}
It is noticed that the inversion about mean operator has the structural symmetry and a mere $2\times2$ matrix, $\Delta_{reduced} = \begin{pmatrix} \frac{2}{N} - 1 & \frac{2}{N} \\ \frac{2}{N} & \frac{2}{N} - 1\end{pmatrix}$, would be sufficient to represent it and operate on a quantum state. Here, each diagonal element would carry the matrix element $\frac{2}{N}-1$ and off diagonal element $\frac{2}{N}$, where $N=2^{n}$ is the dimension. When $\Delta_{reduced}$ operates on each of the elements of the state vector, previously operated by the oracle, only the marked state amplitude will become prominent with the progress of Grover's iterator. 

The {\em Diffusion operator}, when applied to an arbitrary state $|\psi\rangle = \sum_{x=0}^{N-1}c_{x}|x\rangle$, will result in $\sum_{x=0}^{N-1}[-c_{x} + 2\langle c\rangle|x\rangle$. The following Algorithm \ref{alg:invmean} outlines the implementation.

\begin{algorithm}[H]
  \caption{n-qubit Inversion about Mean}
  \label{alg:invmean}
  \SetAlgoLined
  \SetKwFunction{FMain}{inversion\_mean}
  \SetKwProg{Fn}{Function}{:}{end}
  \Fn{\FMain{state: $\psi[N]$, solution state: $|i\rangle$}}{
  $real_sum = imag_sum = 0$\;
  \For {$basis\_state=0;\ basis\_state < N-1;\ basis\_state++$}{
    \If {$\psi[basis\_state] == |i\rangle$ }{
        $real\_sum += \psi[basis\_state].real$\;
        $imag\_sum += \psi[basis\_state].imag$\;
        $n++$\;
    }
  }

  $real\_sum = ((real\_sum * 2) / (1 << n))$\;
  $imag\_sum = ((imag\_sum * 2) / (1 << n))$\;

  \For {$basis\_state=0;\ basis\_state < N-1;\ basis\_state++$}{
    \If {$\psi[basis\_state] == |i\rangle$ }{
        $\psi[basis\_state].real = real\_sum - \psi[basis\_state].real$\;
        $\psi[basis\_state].imag = imag\_sum - \psi[basis\_state].imag$\;
    }
  }
  return 0\;
}
\end{algorithm}

The run-time in this case is again O($N$), compared to O($[2n+f_{c}(n)]N$) in conventional Grover simulator.

\subsection{Simulation of Grover's Search}
The complete Grover search simulation using the proposed approach is outlined in Algorithm \ref{alg:grover1}.

\begin{algorithm}[H]
  \caption{$n$ qubit Grover's Search}
  \label{alg:grover1}
  \SetAlgoLined
  \SetKwFunction{FMain}{grover\_search}
  \SetKwProg{Fn}{Function}{:}{end}
  \Fn{\FMain{search state: $|i\rangle$}}{
  $\psi[N] = |0\rangle^{n}$\;
  $H^{\otimes n}(\psi[N])$\; \tcp*{prepare superposition state}

  \For {$iter=0;\ iter < \frac{\pi * \sqrt{N}}{4};\ iter++$}{
      $oracle(\psi[N],i)$ \tcp*{phase inversion operator on state $|i\rangle$}
      $inversion\_mean(\psi[N])$ \tcp*{apply Inversion about mean operator on state $|\psi\rangle$}      
  }
  return 0\;
  }
\end{algorithm}
The overall time complexity is $O(N^{\frac{1}{2}}N) = O(N^{\frac{3}{2}})$ compared to O($[f_{o}(n)+f_{c}(n)+2n]N^{\frac{3}{2}}$) for a quantum simulator. There is a significant improvement in the time complexity due to optimized computation, which is crucial for general purpose classical computers with limited CPU. The reduction in time complexity by a factor $f_{o}(n)+f_{c}(n)+2n$ is observed to be significant for simulation involving more than $10$ qubits.

\section{Simulation Results}
\subsection{Measurement outcome}
The measurement statistics for a $3$-qubit search problem generated from the proposed simulation, along with the one obtained from IBM Q Experience \cite{ibmbluemix} simulation, for reference, are captured in Fig \ref{grsrch} below.

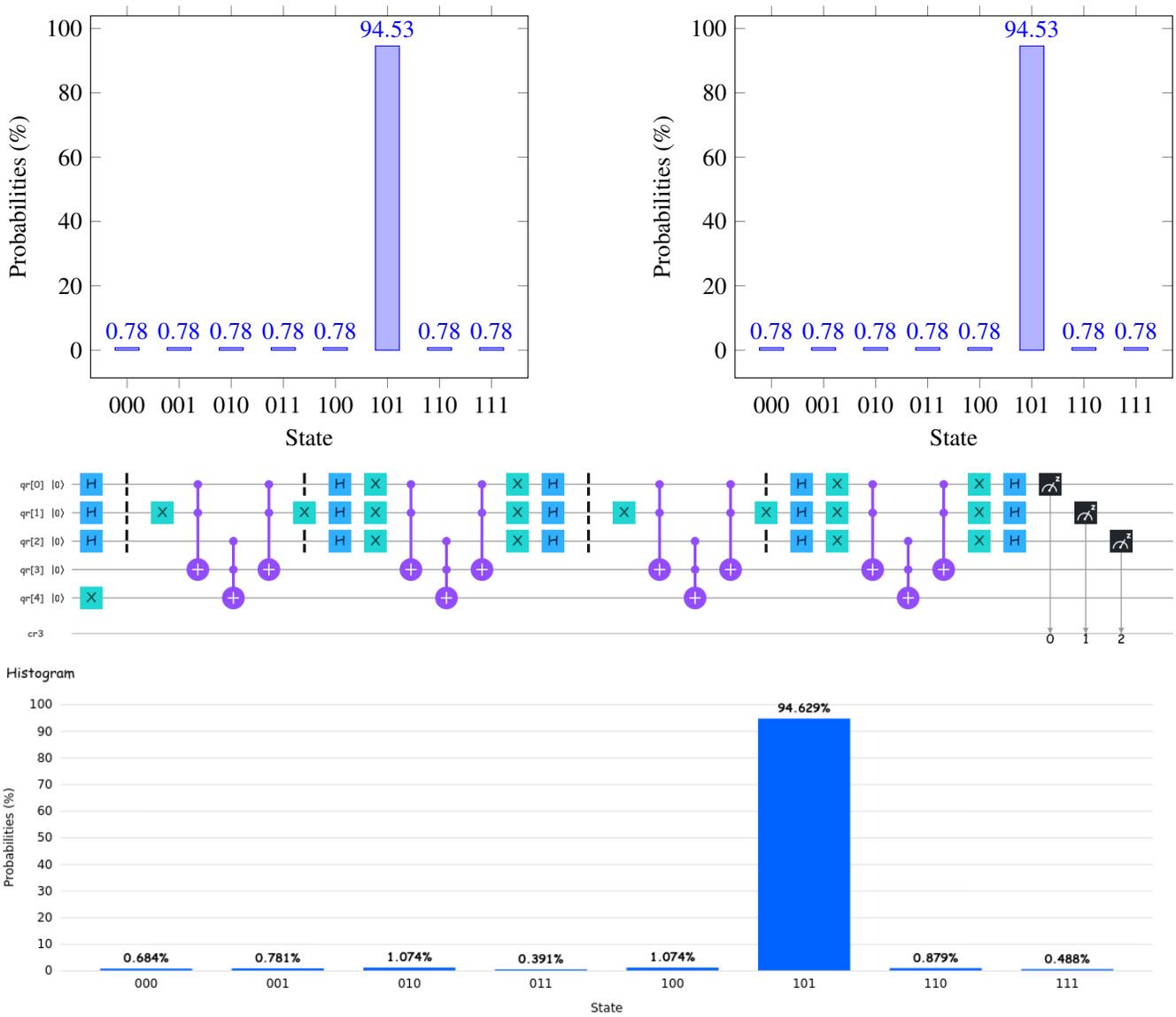
\begin{figure}[!ht]
\centering
\begin{minipage}[b]{.45\textwidth}
 \begin{adjustbox}{width=\linewidth}
  \begin{tikzpicture}
    \begin{axis}[
        ybar,
        ylabel={Probabilities (\%)},
        xlabel={State},
        enlargelimits=0.10,
        symbolic x coords={000,001,010,011,100,101,110,111},
        xtick=data,
        nodes near coords,
        nodes near coords align={vertical},
    ]
    \addplot coordinates { (000, .781) (001, .781) (010, .781) (011, .781) (100, .781) (101, 94.531) (110, .781) (111, .781) };
   \end{axis}
  \end{tikzpicture}
  \end{adjustbox}
 \caption{Three qubits Grover search outcome: solution state $|101\rangle$. (1) Approach 1, (2) Approach 2, (3) IBM Q Simulator} \label{grsrch}
\end{minipage}\hfill
\begin{minipage}[b]{.45\textwidth}
 \begin{adjustbox}{width=\linewidth}
 \begin{tikzpicture}
    \begin{axis}[
        ybar,
        ylabel={Probabilities (\%)},
        xlabel={State},
        enlargelimits=0.1,
        symbolic x coords={000,001,010,011,100,101,110,111},
        xtick=data,
        nodes near coords,
        nodes near coords align={vertical},
    ]
    \addplot coordinates { (000, .781) (001, .781) (010, .781) (011, .781) (100, .781) (101, 94.531) (110, .781) (111, .781) };
   \end{axis}
 \end{tikzpicture}
 \end{adjustbox}
\end{minipage}\hfill

\begin{minipage}[b]{1.0\textwidth}
  \centering
 \begin{adjustbox}{width=\linewidth}
   \includegraphics[scale=.10]{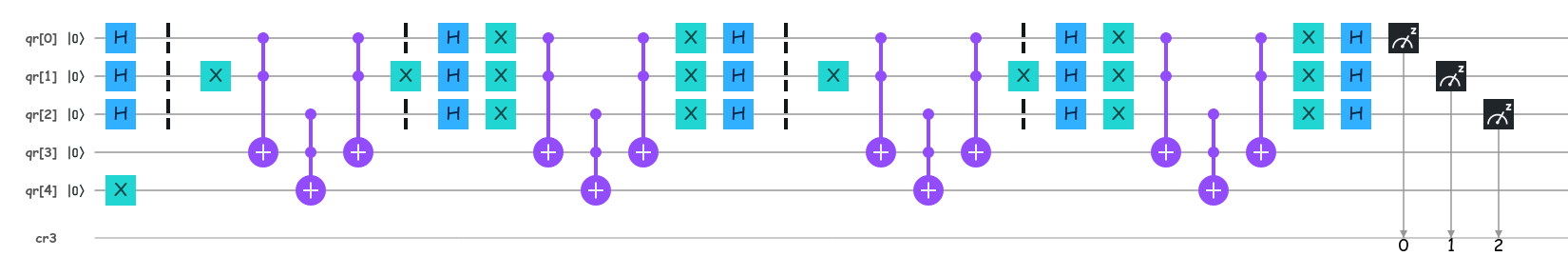}
 \end{adjustbox}
 \end{minipage}\hfill
 \begin{minipage}[b]{1.0\textwidth}
  \centering
 \begin{adjustbox}{width=\linewidth}
  \includegraphics[scale=.10]{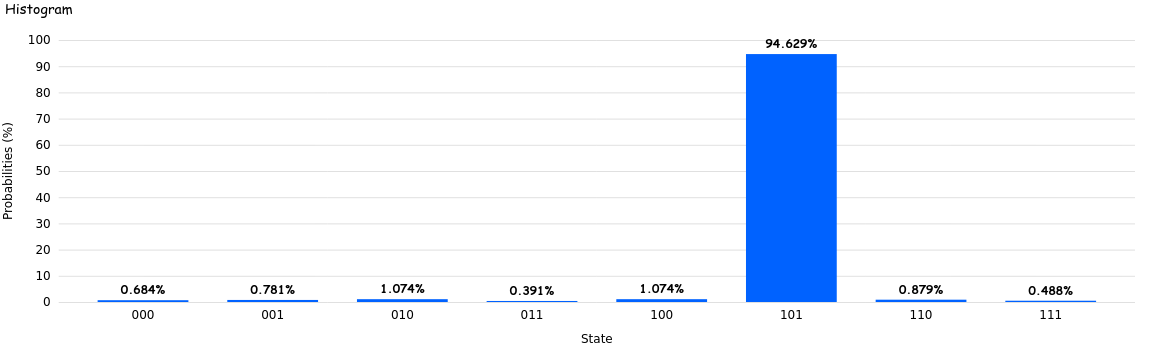}
 \end{adjustbox}
\end{minipage}\hfill
\end{figure}
\newpage

\subsection{Simulator Performance}
The simulations were performed on a laptop with Intel(R) Core(TM) i3-5005U CPU @ 2.00GHz and 8GB RAM. The program was compiled with $-O4$ compiler option and executed on a single CPU/thread. We obtained the performance baseline with the conventional simulation method captured in the background section, which is found to be inline with the existing simulators. This, along with the outcome observed from the proposed simulation approach, is captured in Table \ref{tab:perf}.

\begin{table}[!ht]
  \begin{center}
    \begin{tabular}{|l|l|l|}
      \hline
      Grover Search & \# qubits & minutes:seconds (approx)\\
      \hline
       \multirow{4}{*}{Conventional Simulator} & 10 & 0:0.010 \\
       \cline{2-3}
       & 16 & 0:3.768 \\
       \cline{2-3}
       & 18 & 0:37.708 \\
       \cline{2-3}
       & 20 & 5:57.029 \\
       \cline{2-3}
       \hline
       \multirow{4}{*}{Proposed Simulator} & 10 & 0:0.025\\
       \cline{2-3}
       & 16 & 0:0.349 \\
       \cline{2-3}
       & 18 & 0:2.579 \\
       \cline{2-3}
       & 20 & 0:20.005 \\
       \cline{2-3}
      \hline       
    \end{tabular}
  \end{center}
    \caption{Running time comparison}
    \label{tab:perf}
\end{table}

\newpage
\section{Discussion}
We have shown that the simulation results from the proposed method indicate significant performance improvement, compared to conventional quantum simulators. Though the implementation approach is classical, it is mathematically equivalent to the conventional method. The performance obtained can be further improved multiple folds on high performance classical computers by using the full potential of multi-core CPUs and distributed architecture.

\section*{Acknowledgements}
We would like to thank our mentor Dr. Rajendra K. Bera for giving us an opportunity to be involved in this project, which is part of an ongoing research work towards building a quantum computing framework.

\end{document}